\documentclass[a4]{article}

\usepackage[dvips]{graphics}
\usepackage{epsfig}

\usepackage{amsmath, amsthm, amsfonts, amssymb, bbm}
\usepackage[mathscr]{eucal}

\theoremstyle{plain}
\newtheorem{proposition}{Proposition}[section]
\newtheorem{theorem}[proposition]{Theorem}
\newtheorem{lemma}[proposition]{Lemma}

\theoremstyle{definition}
\newtheorem{definition}[proposition]{Definition}

\theoremstyle{remark}
\newtheorem{remark}[proposition]{Remark}

\newcommand{\ud}{\mathrm{d}}
\newcommand{\del}{\partial}

\newcommand{\pv}[1]{\mathrm{P} \frac{1}{#1}}

\newcommand{\tr}{\mathrm{tr}}

\newcommand{\V}[1]{\mathbf{#1}}

\newcommand{\order}{\mathcal{O}}

\newcommand{\R}{\mathbb{R}}

\newcommand{\betrag}[1]{\left| #1 \right|}
\newcommand{\norm}[1]{\left\| #1 \right\|}

\newcommand{\WDp}[1]{\colon \negthickspace #1 \! \colon \negthickspace }

\renewcommand{\slash}[2][4]{\ensuremath{\rlap{\raisebox{1pt}{$\mspace{#1mu}/$}}#2}}

\newcommand{\dslash}{\slash[2]{\partial}}
\newcommand{\kslash}{\slash[2]{k}}

\newcommand{\lslash}{\slash[0]{l}}

\newcommand{\Vmat}[1]{\mathbf{#1}}
\newcommand{\Sym}{\text{Sym}}
\newcommand{\grad}{\nabla}

\begin{document}



\begin{flushright}
DESY 06-060 \\
ZMP-HH/06-07
\end{flushright}
\hfill
\begin{center}
\setlength{\baselineskip}{20pt}
{\LARGE Dispersion relations in the noncommutative $\phi^3$ and Wess-Zumino model in the Yang-Feldman formalism} \\
\hfill \\
\setlength{\baselineskip}{11pt}
{\large Claus D\"oscher and Jochen Zahn \\ II. Institut f\"ur Theoretische Physik, Universit\"at Hamburg \\ Luruper Chaussee 149, 22761 Hamburg, Germany \\ claus.doescher, jochen.zahn@desy.de \\
\hfill \\
October 13, 2007 \\}
\hfill \\
\end{center}


\begin{center}
{\bf Abstract}
\end{center}

We study dispersion relations in the noncommutative $\phi^3$ and Wess--Zumino model in the Yang--Feldman formalism
at one--loop order. Nonplanar graphs lead to a distortion of the dispersion relation. We find that the strength of this effect
is moderate if the scale of noncommutativity is identified with the Planck scale and parameters typical for a Higgs
field are employed. The contribution of the nonplanar graphs is calculated rigorously using the framework of
oscillatory integrals.



\section{Introduction}

We discuss dispersion relations for quantum field theories on the noncommutative Minkowski space, which is
generated by coordinates $q^{\mu}$ subject to the commutation relations
\begin{equation*}
  [q^{\mu}, q^{\nu}] = i \sigma^{\mu \nu}.
\end{equation*}
Here $\sigma$ is an antisymmetric matrix. Such commutation relations are motivated from Gedanken experiments on
limitations of the localization of experiments~\cite{DFR}. They are also obtained as a limit of open string
theory in the presence of a constant background $B$--field~\cite{SW}. We emphasize that for the space--time uncertainty relations derived in~\cite{DFR} it is crucial that $\sigma$ is nondegenerate, in particular $\sigma^{0i} \neq 0$, i.e., one has space/time noncommutativity. Thus, we focus on this case. We remark that such a $\sigma$ can not be obtained as a limit of string theory~\cite{EField}.

There are several inequivalent approaches to quantum field theory on the noncommutative Minkowski space (NCQFT).
In the modified Feynman rules originally proposed in~\cite{Filk} for both the noncommutative Euclidean and the
Minkowski space, one simply attaches a phase factor depending on the momenta, the so--called twisting, to each
vertex. In cases where the twistings do not cancel, one speaks of a non-planar diagram. Then an oscillating
phase remains in the loop integral. It is part of the folklore of NCQFT that this makes the loop integral
convergent. However, to the best of our knowledge, the precise meaning of these integrals has never been stated.
They are not absolutely convergent and are, with the exception of the tadpole, no Fourier transformations. It is
one of the goals of this paper to give a precise definition for such integrals. Furthermore, to the best of our
knowledge, all calculations in this approach were done in the Euclidean setting. However, since there is no
Osterwalder--Schrader theorem for field theories on the noncommutative Minkowski space, the relation between
calculations in the Euclidean and the Lorentzian metric is obscure in the case of space/time
noncommutativity. In fact there are hints that if such a relation exists at all, it must be quite
complicated~\cite[p.84f]{DorosDiss}.

If one accepts the formal nature of the loop calculations and the transition to the Euclidean signature, the
picture is as follows: If $k$ is the outer momentum of a nonplanar loop, one can argue heuristically that an
original $f(\Lambda)$--divergence, where $\Lambda$ is the UV cutoff, becomes regularized to $f(\betrag{(k \sigma)^2}^{-\frac{1}{2}})$. Thus, a
UV--divergence becomes an IR--divergence. This is the so--called UV--IR mixing first discussed in \cite{NCPertDyn}.
In the case of space/time noncommutativity this approach leads to a violation of unitarity~\cite{GomisMehen}.

The Hamiltonian approach
~\cite{DFR, Sibold} leads to a unitary theory also in the case of space/time noncommutativity. In some cases
these theories are UV--finite~\cite{UVfinite, DoroAvHam}. However, in the case of space/time noncommutativity,
the interacting field does, at tree level, not fulfill the classical equations of motion \cite{DorosDiss, Heslop}.
In the case of electrodynamics, this leads to a violation of the Ward identity~\cite{Ohl}\footnote{In \cite{Heslop}, a different time--ordering, with respect to light--cone coordinates was proposed. While Feynman rules can be formulated quite elegantly in this setting, actual computations seem to be rather involved.}.

Another proposal is to consider Euclidean self-dual theories in the sense of~\cite{Langmann} by adding a
confining potential. In this approach the renormalizability of the $\phi^4$--model has been shown to all
orders~\cite{GrosseWulkenhaar}. However, there is no indication that these models are related to NCQFT on Minkowski spacetime.

Thus, the most promising approach to NCQFT in the case of space/time noncommutativity is the
Yang--Feldman approach~\cite{YF}. It can also be employed in situations where a Hamiltonian quantization is problematic. In particular, it was used in the context of nonlocal field theories, see, e.g., \cite{Moller, Marnelius}. In the context of NCQFT, it was first proposed in \cite{BDFP}. Here the UV-IR mixing manifests itself as a distortion of the dispersion
relation in the infrared. In the case of the $\phi^4$--model, this effect has been shown to be very
strong~\cite{Quasiplanar}. This is to be expected, since the underlying UV--divergence is quadratic. Thus, it is
natural to ask wether the effects are weaker in theories that are only logarithmically divergent\footnote{One
has to bear in mind that it is not clear if the usual power counting arguments can be applied in the
Yang--Feldman approach, in particular in the presence of twisting factors. This will become clearer in Section~\ref{sec:phi3}.}. This is the aim of the present paper
where we consider the $\phi^3$ and the Wess--Zumino model at the one-loop level. It turns out that the effect is
indeed quite weak if one uses the Planck scale as the scale of noncommutativity and uses parameters typical for
a Higgs field. The contributions of the nonplanar graphs, which are made finite by an oscillating factor, are
treated in a rigorous way by the use of the theory of oscillatory integrals~\cite{ReedSimon}. To our knowledge
this has not been done before.

A remark on the issue of Lorentz invariance is in order here. We will see that the self--energy for an outer
momentum $k$ is of the form $\Sigma(k^2, (k \sigma)^2)$. It is thus invariant under Lorentz transformations if
$\sigma$ transforms as a tensor, as has been proposed in~\cite{DFR}. The group velocity, however, should be
computed for fixed $\sigma$. Thus, the dispersion relation can be distorted even though the theory is invariant
under a boost of the reference frame\footnote{See also the discussion in~\cite{WulkenhaarLorentz}, in particular
the distinction between observer and particle Lorentz transformations.}. In the same context, one should remark
that we do not use the concept of twisted Poincar\'e invariance~\cite{Twist} here. 

The noncommutative $\phi^3$-model has already been treated in~\cite{NCPertDyn, Raamsdonk} in the context of the
modified Feynman rules, in~\cite{DoroAvHam} in a Hamiltonian setting, and in~\cite{Grosse} in
the Euclidean self--dual setting.



The noncommutative Wess--Zumino model was first discussed in~\cite{Rivelles} for space/space noncommutativity in
the setting of the modified Feynman rules. It was shown that the UV--IR mixing is much weaker as in the
$\phi^4$--theory, so that the the theory is renormalizable to all orders.


The paper is organized as follows: In Section~\ref{sec:DispRel} we discuss how to compute momentum-dependent
mass and field strength renormalization in the Yang--Feldman approach and to extract the corresponding group
velocity. In Section~\ref{sec:phi3} we apply this machinery to the noncommutative $\phi^3$--model at second
order, i.e., for one loop. In particular, we compute the distortion of the group velocity for parameters typical
for a Higgs field. In Section~\ref{sec:WZ} we treat the noncommutative Wess--Zumino model, also at one--loop
order. We show and discuss the fact that the local SUSY current is not conserved in the interacting case. We also
compute the momentum dependent mass and field strength normalization and show that the distortion of the group
velocity is simply twice that of the $\phi^3$--case. The oscillating integrals so far have only been calculated
formally. A rigorous calculation in the sense of oscillatory integrals is presented in Section~\ref{sec:OscInt}. It turns out that the formal results are indeed correct.
We conclude with a summary and an outlook.

\section{Dispersion Relations in the Yang--Feldman formalism}
\label{sec:DispRel}

We want to discuss how to compute (possibly momentum dependent) mass and field strength renormalizations in the Yang-Feldman formalism. In this formalism, the interacting field is recursively defined as a formal power series in the coupling constant. As an example, we consider a commutative scalar theory and a localized mass term as interaction, i.e., we have the equation of motion
\begin{equation*}
  (\Box + m^2) \phi(x) = - \bar m^2 g(x) \phi(x), 
\end{equation*}
where $g$ is a test function. Making the ansatz
\begin{equation*}
  \phi = \sum_{n=0}^\infty \bar m^{2n} \phi_n
\end{equation*}
for the interacting field, this leads to the equations
\begin{align*}
  (\Box + m^2) \phi_0 & = 0, \\
  (\Box + m^2) \phi_n & = - g \phi_{n-1}, \quad n \geq 1.
\end{align*}
Obviously, $\phi_0$ is a free field. We identify it with the incoming field. Then the higher order terms are given recursively by
\begin{equation*}
  \phi_n = \Delta_R \times (g \phi_{n-1}), \quad n \geq 1,
\end{equation*}
where $\times$ denotes the convolution and $\Delta_R$ the retarded propagator at mass $m$. We define the observable
\begin{equation}
\label{eq:phi_f}
  \phi(f) = \int \ud^4x \ f(x) \phi(x) = \int \ud^4k \ \hat f(-k) \hat \phi(k),
\end{equation}
where the hat denotes the Fourier transform. We are now interested in the Wightman two--point function
\begin{equation}
\label{eq:2pt}
  \langle \phi(f) \phi(h) \rangle
\end{equation}
of the interacting field. The vacuum state here is the vacuum state for the free field $\phi_0$, i.e., in order to compute the above, one has to express $\phi$ solely in terms of $\phi_0$ and then determine the vacuum expectation value. At zeroth order in $\bar m^2$, we obtain the usual free two--point function
\begin{equation}
\label{eq:2pt0order}
  \langle \phi_0(f) \phi_0(h) \rangle = (2 \pi)^2  \int \ud^4k \ \hat{f}(-k) \hat{h}(k) \hat{\Delta}_+(k).
\end{equation}
At first order in $\bar{m}^2$, we get
\begin{multline*}
   \langle \phi_{1}(f) \phi_{0}(h) \rangle + \langle \phi_{0}(f) \phi_{1}(h) \rangle = \\
   -  (2 \pi)^2 \int \prod_{i=0}^1 \ud^4k_i \ \hat{f}(-k_0) \hat{h}(k_1) \hat{g}(k_0-k_1)  \left\{ \hat{\Delta}_{R}(k_0) \hat{\Delta}_+(k_1) + \hat{\Delta}_{+}(k_0) \hat{\Delta}_A(k_1) \right\}.
\end{multline*}
Here $\Delta_A$ is the advanced propagator. It has been shown in \cite{AdLim} that, under quite general assumptions, in the adiabatic limit $g \to 1$, i.e.,
$\hat{g} \to (2\pi)^2 \delta$, this becomes
\begin{equation}
\label{eq:1stOrder}
 - 2 \pi \int \ud^4k \ \hat{f}(-k) \hat{h}(k) \theta(k^0) \delta'(k^2-m^2).
\end{equation}
Obviously, this can be interpreted as the first order term in an expansion of $\Delta_+(m^2+\bar{m}^2, \cdot)$
around $m^2$.

When considering noncommutative field theories, the following changes have to be made: Fields and test functions are now functions of the noncommuting coordinates $q^\mu$, so that products are given by
\begin{align}
  f(q) h(q) & = (2\pi)^{-4} \int \ud^4k \ud^4l \ \hat f(k) \hat h(l) e^{-ikq} e^{-ilq} \nonumber\\
  \label{eq:product}
                & = \int \ud^4k \ e^{-ikq} \int \ud^4l \ \hat f(k-l) \hat h(l) e^{\frac{i}{2} k \sigma l}. 
\end{align}
Here $\hat f$ denotes the Fourier transform of the Weyl symbol of $f(q)$. Alternatively, one could use functions of $x$ and the Weyl--Moyal $\star$--product. The integral (trace) is defined as usual as
\begin{equation*}
  \int \ud^4q \ f(q) = (2\pi)^2 \hat f(0).
\end{equation*}
Then, analogously to (\ref{eq:phi_f}), we have
\begin{equation*}
  \phi(f) = \int \ud^4q \ f(q) \phi(q) = \int \ud^4k \ \hat f(-k) \hat \phi(k).
\end{equation*}
The Yang--Feldman series can be set up exactly as before, i.e., $\phi_0$ is the free field and for $n \geq 1$, we have\footnote{Here the infrared cutoff was implemented by multiplying the ``interaction term'' $\bar m^2 \phi(q)$ in the equation of motion with a ``test function'' $g(q)$ from the left. One can also use more symmetric products, for details see~\cite{AdLim}.}
\begin{align*}
  \phi_n(q) & = \int \ud^4x \ \Delta_R(x) g(q-x) \phi_{n-1}(q-x) \\
                 & = (2 \pi)^{-2} \int \ud^4k \ \hat \Delta_R(k) e^{-ikq} \int \ud^4l \ \hat g(k-l) \hat \phi_{n-1}(l) e^{\frac{i}{2} k \sigma l}.
\end{align*}
It was shown in \cite{AdLim} that also in this case one obtains (\ref{eq:1stOrder}) as the first order contribution to the two--point function in the adiabatic limit $\hat g(k) \to (2 \pi)^2 \delta(k)$.

\subsection{Interactions}

Now we consider truly interacting models. For simplicity we start with a scalar field theory on the ordinary Minkowski space. The coupling constant is denoted by $\lambda$. When computing the two--point function (\ref{eq:2pt}), one finds again (\ref{eq:2pt0order}) as the zeroth order contribution.  In the models discussed in this paper, there is no $\order (\lambda)$ contribution. At second order, one finds the three terms
\begin{equation}
\label{eq:3terms}
   \langle \phi_2(f) \phi_0(h) \rangle + \langle \phi_0(f) \phi_2(h) \rangle + \langle \phi_1(f) \phi_1(h) \rangle.
\end{equation}
As we will see later, the third term is a contribution to the continuous spectrum and thus not interesting at the moment. In order to treat the first two terms, we notice that in the models discussed here, $\phi_2$ is formally of the form
\begin{equation}
\label{eq:phi_2}
  \phi_2 = (2\pi)^{-2} \Delta_R \times ( g ( \check \Sigma \times (g \phi_0 ))) + \text{ n.o.},
\end{equation}
where n.o. stands for a term that is normal ordered and whose spectrum has no overlap with the positive or negative mass shell if the support of $\hat g$ is chosen small enough. 
Thus, this term drops out in the first two terms in (\ref{eq:3terms}). The $\Sigma$ in the first term will in general be divergent and has to be renormalized, which we assume in the following. Then the first term in (\ref{eq:phi_2}) is quite similar to $\phi_1$ in the case of a mass term as interaction. It is thus not very surprising that, using the same techniques as in \cite{AdLim}, one can show (for details see \cite{Claus, InPrep}) that in the adiabatic limit $g \to 1$, one obtains
\begin{equation}
\label{eq:2pt2ndOrder}
  - (2\pi)^2 \int \ud^4k \ \hat f(-k) \hat h(k) \Sigma(k) \tfrac{\del}{\del m^2} \hat \Delta_+(k),
\end{equation}
for the first two terms in (\ref{eq:3terms}) under the condition that $\Sigma(k) = \Sigma(-k)$ in a neighborhood of the mass shell. Here $\Sigma$ is the Fourier transform of $\check \Sigma$ and can be identified with the self--energy. In the commutative case, $\Sigma(k)$ is only a function of $k^2$, and (\ref{eq:2pt2ndOrder}) corresponds to a mass and field strength renormalization
\begin{align*}
  \delta m^2 & = - \lambda^2 \Sigma(m^2), \\
  \delta Z & = - \lambda^2 \tfrac{\del}{\del k^2} \Sigma(m^2).
\end{align*}

In the noncommutative case, a rigorous adiabatic limit is not possible because of UV-IR mixing effects (for details, see~\cite{Claus, InPrep}). We thus take a pragmatic point of view and work formally, i.e., without infrared cutoff. In analogy to (\ref{eq:phi_2}), we write $\phi_2$ in the form
\begin{equation*}
  \hat \phi_2(k) = (2 \pi)^2 \hat \Delta_R(k) \Sigma(k) \hat \phi_0(k) + \text{ n.o.}
\end{equation*}
and take this as an implicit definition of $\Sigma$ (again, we assume $\Sigma$ to be renormalized). If then $\Sigma(k) = \Sigma(-k)$ in a neighborhood of the mass shell, we use (\ref{eq:2pt2ndOrder}) as the sum of the first two terms in (\ref{eq:3terms}). Now $\Sigma(k)$ is in general not only a function of $k^2$, but also of $(k \sigma)^2$. Thus, we obtain momentum-dependent mass and field strength renormalizations:
\begin{align}
\label{eq:M}
  \delta m^2((k \sigma)^2) & = - \lambda^2 \Sigma(m^2, (k \sigma)^2), \\
\label{eq:Z}
  \delta Z((k \sigma)^2) & = - \lambda^2 \tfrac{\del}{\del k^2} \Sigma(k^2, (k \sigma)^2) |_{k^2=m^2}.
\end{align}

\begin{remark}
Although the naming might suggest that these terms should be subtracted, we do not do so, since they are neither
local, nor, in general, divergent. We remark, however, that such a subtraction has been proposed in~\cite{LiaoSibold}.
\end{remark}



\subsection{The group velocity}
\label{sec:GroupVelocity}

The sum of the zeroth order term (\ref{eq:2pt0order}) and the second order contribution (\ref{eq:2pt2ndOrder}) can be interpreted as the expansion (in $\lambda$) of
\begin{equation}
\label{eq:2ptNC}
 2\pi \int \ud^4k \ \hat{f}(-k) \hat{h}(k) \theta(k^0) \delta(k^2-m^2+ \lambda^2 \Sigma(k^2, (k \sigma)^2)) + \order (\lambda^4).
\end{equation}
This can be interpreted as a change of the dispersion relation.
\begin{remark}
\label{rem:LSZ}
This modification of the dispersion relation is a manifestation of the breaking of particle Lorentz invariance, cf. the discussion in the introduction. However, particle Lorentz invariance of the asymptotic fields is a crucial ingredient of scattering theory and the LSZ relations, which are part of the foundations of quantum field theory. In this sense, the conceptual basis of the present approach is rather shaky. In the following, we will take a phenomenological standpoint and compute the distortion of the dispersion relation for different models in order to check if they are realistic. 
\end{remark}
We now discuss how to extract the group velocity in the above setting. From~(\ref{eq:2ptNC}), and allowing for a finite local mass and field strength renormalization, we get the dispersion relation
\begin{equation}
\label{eq:ImplicitDispRel}
 F(k) = k^2 - m^2 + \lambda^2 \left( \Sigma(k^2, (k \sigma)^2) - \alpha + \beta k^2 \right) + \order(\lambda^4) = 0.
\end{equation}
For a given spatial momentum $\V{k}$ we want to compute the corresponding $k^0$ that solves
(\ref{eq:ImplicitDispRel}) as a formal power series in $\lambda$. We find
\begin{equation}
\label{eq:k0}
  k^0 = \omega_k - \lambda^2 \frac{1}{2 \omega_k} \left( \Sigma(m^2, (k_+ \sigma)^2) - \alpha + \beta m^2 \right) + \order(\lambda^4).
\end{equation}
Note that in $\omega_k = \sqrt{\betrag{\V{k}}^2 + m^2}$ and $k_+=(\omega_k, \V{k})$ the bare mass $m$ enters.
The group velocity is then given by
\begin{multline*}
  \nabla k^0 = \frac{\V{k}}{\omega_k} + \lambda^2 \frac{\V{k}}{2 \omega_k^3} \left( \Sigma(m^2, (k_+ \sigma)^2) - \alpha + \beta m^2 \right) \\ - \lambda^2 \frac{1}{2 \omega_k} \nabla (k_+ \sigma)^2 \frac{\del}{\del (k \sigma)^2} \Sigma(m^2, (k_+ \sigma)^2) + \order(\lambda^4).
\end{multline*}
By comparison with (\ref{eq:k0}), we get
\begin{equation*}
  \nabla k^0 = \frac{\V{k}}{k^0} - \lambda^2 \frac{\nabla (k_+ \sigma)^2}{2 k^0}  \frac{\del}{\del (k \sigma)^2} \Sigma(m^2, (k_+ \sigma)^2) + \order(\lambda^4).
\end{equation*}
In order to make things more concrete, we choose a particular $\sigma$, namely,
\begin{equation}
\label{eq:sigma_0}
  \sigma = \sigma_0 = \lambda_{nc}^2 \begin{pmatrix} 0 & - \mathbbm{1} \\ \mathbbm{1} & 0 \end{pmatrix}.
\end{equation}
Then we have
\begin{equation}
\label{eq:k_sigma_2}
 (k \sigma_0)^2 = - \lambda_{nc}^{4} \left( k^2 + 2 \betrag{\V{k_{\bot}}}^2 \right)
\end{equation}
with $\V{k_{\bot}} = (k_1, 0, k_3)$. We also define $\V{k_{||}}=(0,k_2,0)$. Thus, in the case $\sigma =
\sigma_0$, we find
\begin{equation}
\label{eq:GroupVelocity}
  \V{\nabla} k^0 = \frac{\V{k_{||}}}{k^0} + \frac{\V{k_{\bot}}}{k^0} \left( 1 +  2 \lambda^2 \lambda_{nc}^4 \frac{\del}{\del (k \sigma)^2} \Sigma(m^2, (k_+ \sigma_0)^2) \right) + \order(\lambda^4).
\end{equation}
\begin{remark}
This treatment differs slightly from the one given in~\cite{Quasiplanar}. There, $\Sigma$ is not Taylor expanded in $\lambda$. Then the argument of $\Sigma$ in (\ref{eq:GroupVelocity}) is not restricted to the mass $m$ shell. It follows that by tuning $\alpha$ and $\beta$ one can make the deviation arbitrarily small, which is not possible here.
\end{remark}

\begin{remark}
\label{rem:FSRen}
The modification of the dispersion relation can be interpreted as an effect of the momentum--dependent mass renormalization~(\ref{eq:M}), since $\lambda^2 \Sigma$ in (\ref{eq:GroupVelocity}) can be replaced by $- \delta m^2$. The momentum--dependent field strength renormalization (\ref{eq:Z}), on the other hand, multiplies, in momentum space, the free propagators, in particular the retarded propagator. In position space, this can be interpreted as a smearing of the source, and thus as a non--local effect. In~\cite{InPrep}, this is explained in more detail, and the effect is computed for the case of noncommutative supersymmetric electrodynamics. In particular, it is shown that, surprisingly, the range of this nonlocality is independent from the scale of noncommutativity.
\end{remark}

\section{The $\phi^3$--model}
\label{sec:phi3}

We now apply the above tools to the noncommutative $\phi^3$--model and compute the momentum--dependent mass and field strength
renormalization and the distortion of the group velocity at second order. We start from the equation of motion
\begin{equation*}
  ( \Box + m^2 ) \phi = \lambda \phi^2.
\end{equation*}
The Yang--Feldman ansatz $\phi = \sum_n \lambda^n \phi_n$, and the identification of $\phi_0$ with the incoming
field then leads to
\begin{align*}
  \phi_1 & = \Delta_R \times ( \phi_0 \phi_0 ), \\
  \phi_2 & = \Delta_R \times ( \phi_1 \phi_0 + \phi_0 \phi_1 ).
\end{align*}
We substract the tadpole from the start, i.e., we use normal ordering and redefine
\begin{equation*}
  \phi_1 = \Delta_R \times ( \WDp{\phi_0 \phi_0} ).
\end{equation*}

Now we want to compute the two--point function of the interacting field. At zeroth order, we find the usual
result (\ref{eq:2pt0order}). At first order, there is no contribution. At second order, there are the three terms (\ref{eq:3terms}).
We first focus on the sum of the first two terms. As discussed in the previous section, we treat it by computing the self--energy $\Sigma(k)$. Performing the contractions in $\phi_2$, we obtain
\begin{align*}
  \hat \phi_2(k) = & (2\pi)^2 \hat \Delta_R(k) \hat \phi_0(k) \\
  & \times \int \ud^4l \ \hat \Delta_R(k-l) \left\{ \hat \Delta_+(-l) \left( 1 + e^{-ik\sigma l} \right) + \hat \Delta_+(l) \left( 1 + e^{ik\sigma l} \right) \right\} \\
  & + \text{ n.o.}
\end{align*}
Thus, $\Sigma$ is given by
\begin{equation*}
  \Sigma(k) =  \int \ud^4l \ \hat \Delta_R(k-l) \left\{ \hat \Delta_+(-l) \left( 1 + e^{-ik\sigma l} \right) + \hat \Delta_+(l) \left( 1 + e^{ik\sigma l} \right) \right\}.
\end{equation*}
This can be split into a planar part not involving the phase factors and a nonplanar part. The planar part is precisely half of the self--energy of the commutative $\phi^3$ model.

For the following consideration, it is important that we are only interested in $\Sigma(k)$ in a small neighborhood of the mass shell. But also the loop momentum $l$ is confined to the mass shell, so if $(m - \epsilon)^2 < k^2 < (m + \epsilon)^2$, then either $(k-l)^2 < \epsilon^2$ or $(k-l)^2>(2m-\epsilon)^2$. Thus, the singularity of $\hat \Delta_R(k-l)$ is not met and the $i \epsilon$--prescription does not matter: One may simply write
\begin{equation*}
  \hat \Delta_R(k-l) = (2 \pi)^{-2} \frac{-1}{(k-l)^2 - m^2} = (2 \pi)^{-2} \frac{-1}{k^2 - 2 k \cdot l}.
\end{equation*}

We begin by discussing the planar part
\begin{equation}
\label{eq:Sigma_pl}
  \Sigma_{pl}(k) =  \int \ud^4l \ \hat \Delta_R(k-l) \left\{ \hat \Delta_+(-l) + \hat \Delta_+(l) \right\}.
\end{equation}
As usual, this expression is not well--defined. Because of the preceding remark, it is straightforward to show that at least formally $\Sigma_{pl}(k) = \Sigma_{pl}(-k)$ in a neighborhood of the mass shell. It has been shown in~\cite{BDFP} that
\begin{equation*}
  \Delta_R \cdot (\Delta_+ + \Delta_-) = \Delta_F^2 - \Delta_-^2
\end{equation*}
holds. Here $\Delta_-^2$ is well--defined, while $\Delta_F^2$ has the usual logarithmic divergence. Alternatively, one may argue with the following formal calculation: Because of Lorentz invariance, we may choose
$k=(k_0, \V{0})$. Then
\begin{align}
  \Sigma_{pl}(k) = & - (2 \pi)^{-3} \int \frac{\ud^3 l}{2 \omega_l} \ \left( \frac{1}{k_0^2 - 2 k_0 \omega_l} + \frac{1}{k_0^2 + 2 k_0 \omega_l} \right) \nonumber \\
\label{eq:F_pl}
  = & - 2 (2 \pi)^{-2} \int_0^{\infty} \ud l \ \frac{l^2}{\omega_l ( k_0^2 - 4 \omega_l^2)},
\end{align}
which diverges logarithmically. We note that it is necessary to consider the sum of the two terms in (\ref{eq:Sigma_pl}). The individual terms are linearly divergent. It is a priori not clear if the same cancellation takes place in the presence of the twisting factors, i.e., in the nonplanar part. Hence, the validity of power counting arguments for noncommutative field theories in the Yang--Feldman formalism is doubtful.

Finally, we remark that the field strength renormalization is finite. Using (\ref{eq:Z}), one computes
\begin{equation}
\label{eq:Z_pl}
  \delta Z= (2 \pi)^{-2} \frac{3 - \frac{2 \pi}{\sqrt{3}}}{12 m^2}.
\end{equation}

\subsection{The nonplanar part}
\label{sec:phi3np}

We now want to discuss the nonplanar part of $\Sigma(k)$, i.e.,
\begin{equation}
\label{eq:F_NP}
  \Sigma_{np}(k) = \int \ud^4l \ \hat{\Delta}_+(l) e^{i k \sigma l} \left( \hat{\Delta}_{R}(k-l) + \hat{\Delta}_{R}(k+l)
  \right),
\end{equation}
for $k$ in a neighborhood of the mass shell. In particular, we want to show that it is finite and that $\Sigma_{np}(k) = \Sigma_{np}(-k)$ there.
Note that the above integral is neither absolutely convergent nor a Fourier transformation (since $k$ does not only appear in the phase factor). In the following, we compute this integral in a formal way. In Section~\ref{sec:OscInt} we show that~(\ref{eq:F_NP}) can be defined as an oscillatory integral and that a calculation in this framework gives the same result as our formal calculation.

First of all we note that if $\Sigma_{np}(k)$ is well defined, then it is invariant under
the Lorentz transformation
\begin{equation*}
  k \to k \Lambda, \qquad \sigma \to \Lambda^{-1} \sigma {\Lambda^T}^{-1}.
\end{equation*}
Thus, instead of computing the above at $k, \sigma$ we may compute it at $k'=k \Lambda, \sigma'= \Lambda^{-1}
\sigma {\Lambda^T}^{-1}$. Since at the one-loop level we are only interested in $\Sigma_{np}(k)$ in a neighborhood of the mass shell,
we may choose $k'=(\sqrt{k^2}, \V{0})$. Since $\sigma'$ is antisymmetric, $k' \sigma'$ has vanishing time component.
We denote its spatial component by $\underline{k' \sigma'}$. Then we have
\begin{align}
  \Sigma_{np}(k) = & - (2 \pi)^{-3} \int \frac{\ud^3 l}{2 \omega_l} \ \left( \frac{e^{-i \underline{k' \sigma'} \cdot \V{l}}}{k^2 - 2 \sqrt{k^2} \omega_l} + \frac{e^{-i \underline{k' \sigma'} \cdot \V{l}}}{k^2 + 2 \sqrt{k^2} \omega_l} \right) \nonumber \\
  = & - 2 (2 \pi)^{-3} \int \frac{\ud^3 l}{2 \omega_l} \ \frac{1}{k^2 - 4 \omega_l^2} \cos (\underline{k' \sigma'} \cdot \V{l}) \nonumber \\
\label{eq:F_np}
  = & - 2 (2 \pi)^{-2} \int_0^{\infty} \ud l \ \frac{l^2}{\omega_l (k^2 - 4 \omega_l^2)} \frac{\sin l \sqrt{-(k \sigma)^2}}{l \sqrt{-(k
  \sigma)^2}}.
\end{align}
In the first step we used the the symmetry properties of the integrand. In the next step we used $(k \sigma)^2 = (k' \sigma')^2 = - \betrag{ (\underline{k' \sigma'}) }^2$.
Obviously, the integral is finite and only a function of $k^2$ and $(k \sigma)^2$. Furthermore, $\Sigma_{np}(k) = \Sigma_{np}(-k)$.

\begin{figure}
\epsfig{file=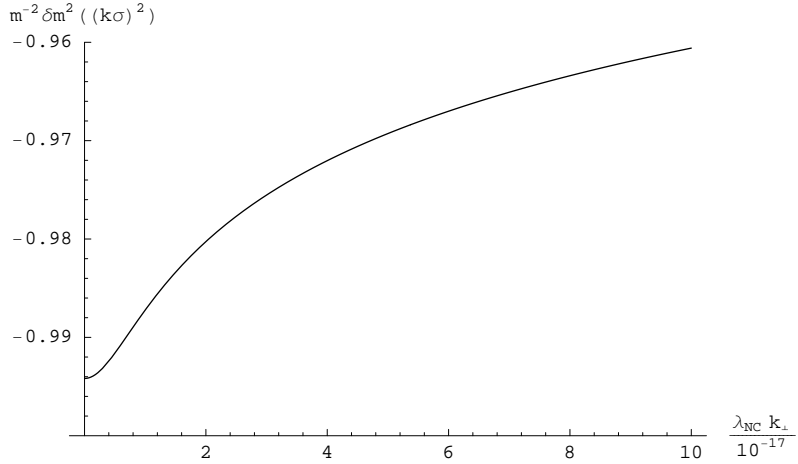,scale=0.75} 
\caption{\label{fig:M}The relative mass correction $m^{-2} \delta m^2((k \sigma)^2)$ as a function of the perpendicular
momentum $k_{\bot}$.}
\end{figure}

In order to estimate the strength of the distortion of the dispersion relation, we calculate $\delta m^2((k \sigma)^2)$ and $\delta Z((k \sigma)^2)$ numerically. We use the parameters $\sigma=\sigma_0$ (cf. (\ref{eq:sigma_0})), $m=10^{-17} \lambda_{nc}^{-1}$ and $\lambda = m$. If $\lambda_{nc}$ is identified with the Planck length, this corresponds to a mass of about $100 \text{GeV}$, i.e., the estimated order of magnitude of the Higgs mass. The chosen value of $\lambda$ is slightly above the expectation for the cubic term in the Higgs potential ($\sim 0.6m$). Figure~\ref{fig:M} shows the relative mass correction $m^{-2} \delta m^2((k \sigma)^2)$ as a function of the perpendicular momentum $k_{\bot}$, obtained with the numerical integration method of {\sc mathematica} (for the definition of $k_\bot$, see Section~\ref{sec:GroupVelocity}). We see that the relative mass shift is of order $1$ for small perpendicular momenta. This might look like a strong effect. However, we have the freedom to apply a finite mass renormalization in order to restore the rest mass. The important question is rather how strong the momentum dependence of the mass renormalization is. As can be estimated from Figure~\ref{fig:M}, it is at the \%-level for perpendicular momenta of the order of the mass. As a consequence, also the distortion of the group velocity is of this order, as we will show below.


The plot for $\delta Z((k \sigma)^2)$ for the same parameters is not very interesting, since $\delta Z$ is constant, $-1.32477 \cdot 10^{-3}$, within machine precision. This coincides with the planar contribution (\ref{eq:Z_pl}). The reason
for this is easily understood: If one differentiates the integrand in (\ref{eq:F_np}) with respect to $k^2$, one obtains a
function that, even without the factor
\begin{equation*}
  \frac{\sin l \sqrt{-(k \sigma)^2}}{l \sqrt{-(k \sigma)^2}},
\end{equation*}
is integrable. Without this factor, it would coincide with the corresponding planar expression obtained by
differentiating~(\ref{eq:F_pl}). But the above factor deviates from 1 appreciably only for $l \sim (-(k
\sigma)^2)^{-\frac{1}{2}}$, i.e., for very high energies, where the rest of the integrand is negligible.



\begin{figure}
\epsfig{file=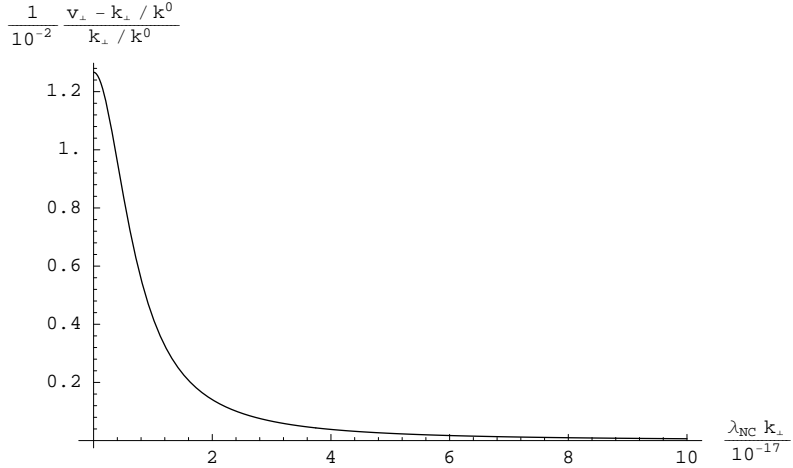,scale=0.75} 
\caption{\label{fig:GroupVel}The distortion of the group velocity in perpendicular direction as a function of
the perpendicular momentum $k_{\bot}$.}
\end{figure}

According to equation (\ref{eq:GroupVelocity}), the deviation of the group velocity from the phase velocity in
the perpendicular direction is, to lowest order in $\lambda$, given by $2 \lambda^2 \lambda_{nc}^4
\frac{\del}{\del (k \sigma)^2} \Sigma_{np}$. Figure~\ref{fig:GroupVel}  shows this quantity for the same parameters
as above.
The deviation is biggest for small perpendicular momenta and at the \%-level.

We see that in the $\phi^3$ model the distortion of the dispersion relation is moderate for realistic masses and couplings. This is in sharp contrast to the situation in the $\phi^4$ model, where realistic dispersion relations could only be obtained for masses close to the noncommutativity scale~\cite{DorosDiss}.

\subsection{The 2--particle spectrum}

We now discuss the third term in (\ref{eq:3terms}).
We obtain
\begin{equation*}
  (2\pi)^4 \int \ud^4 k \hat f(-k) \hat h(k) \hat \Delta_R(k) \hat \Delta_A(k) \left( ( \Delta_+ \cdot \Delta_+)\hat{}(k) + (\Delta_+ \star_{2 \sigma} \Delta_+)\hat{}(k) \right). 
\end{equation*}
Here $\star_{2 \sigma}$ is the $\star$-product at $2 \sigma$, i.e., the product corresponding to the twisting
factor $e^{i k \sigma l}$. Like $\Delta_+ \cdot \Delta_+$, $\Delta_+ \star_{2 \sigma} \Delta_+$ is a well--defined distribution, as can be seen in momentum space. It has its
support above the $2m$ mass shell, thus this term corresponds to the multi--particle spectrum. Using Lorentz invariance
as above, one can compute
\begin{equation*}
  (\Delta_+ \star_{2 \sigma} \Delta_+)\hat{}(k) = \theta(k^2-4m^2) (2 \pi)^{-3} \frac{\sin \left( \sqrt{-(k \sigma)^2} \sqrt{\frac{1}{4} k^2 - m^2} \right)}{2 \sqrt{k^2} \sqrt{-(k \sigma)^2}}.
\end{equation*}
In the limit $(k \sigma)^2 \to 0$, this gives back the commutative result. Note that deviations from the
commutative case become appreciable for $- (k \sigma)^2 \sim k^{-2}$, i.e. if $\sqrt{k^2}$ or the transversal
momentum $k_{\bot}$ is of the order $\frac{\lambda_{nc}^2}{\sqrt{k^2}}$. This is obviously no threat to
phenomenology.


\section{The Wess--Zumino model}
\label{sec:WZ}

In this section we consider the Wess--Zumino model on the noncommutative Minkowski space. We use the standard
supersymmetric noncommutative Minkowski space, in which the (anti-) commutators involving the fermionic
variables $\theta, \bar{\theta}$ are unchanged~\cite{SUSYNCMink}. In order to arrive at the equations of motion
for the component fields, we start from the Lagrangean in superfield form, taking particular care for the order
of the fields in the different terms\footnote{This is important, since for example the tadpole corresponding to
the interaction term $\phi^* \phi \phi^* \phi$ does not have a twisting factor, in contrast to the interaction
term $\phi^* \phi^* \phi \phi$, as has already been noted in~\cite{UVIRNote}.}.

In superfield form the Wess--Zumino model is given by the following Lagrangean\footnote{In the following, we use
the conventions of \cite{WessBagger}, except for the metric, which we choose to have signature $(+ - - -)$.
Accordingly, we also changed the sign of $\sigma^0$, and thus also of $\gamma^0$ and $\gamma^5$.}:
\begin{equation*}
  L = \bar{\Phi} \Phi |_{\theta^2 \bar{\theta}^2} + \left\{ \left( \frac{m}{2} \Phi \Phi + \frac{\lambda}{3} \Phi \Phi \Phi \right) |_{\theta^2} + \text{ h.c.}
  \right\}.
\end{equation*}
Here $\Phi$ is the chiral superfield
\begin{equation*}
  \Phi = \phi + \sqrt{2} \theta \chi + \theta^2 F - i \theta \sigma^{\mu} \bar{\theta} \del_{\mu} \phi + \frac{i}{\sqrt{2}} \theta^2 \del_{\mu} \chi \sigma^{\mu} \bar{\theta} - \frac{1}{4} \theta^2 \bar{\theta}^2 \Box \phi,
\end{equation*}
where $\phi$ and $F$ are complex scalar fields and $\chi$ is a Weyl spinor.
In component fields the action is then, up to surface terms,
\begin{multline*}
  S = \int \ud^4q \  \Big( - i \del_{\mu} \bar{\chi} \bar{\sigma}^{\mu} \chi - \phi^* \Box \phi + F^* F  \\ + \left\{ \left( m \left( \phi F - \tfrac{1}{2} \chi \chi \right) + \lambda \left( \phi \phi F - \chi \chi \phi \right) \right) + \text{ h.c.} \right\} \Big).
\end{multline*}
This leads to the equations of motion
\begin{gather*}
  F + m \phi^* + \lambda \phi^* \phi^* = 0 \\
  - \Box \phi + m F^* + \lambda (\phi^* F^* + F^* \phi^*) - \lambda \bar{\chi} \bar{\chi} = 0 \\
   i \bar{\sigma}^{\mu} \del_{\mu} \chi - m \bar{\chi} - \lambda (\phi^* \bar{\chi} + \bar{\chi} \phi^* ) = 0.
\end{gather*}
We eliminate $F$ using its equation of motion. Furthermore, we introduce the Majorana spinor
\begin{equation*}
  \psi = \frac{1}{\sqrt{2}} \begin{pmatrix} \chi_{\alpha} \\ \bar{\chi}^{\dot{\alpha}} \end{pmatrix}, \quad \bar{\psi} = \psi^\dagger \gamma^0 = \frac{1}{\sqrt{2}}  ( \chi^{\alpha},  \bar{\chi}_{\dot{\alpha}} )
\end{equation*}
and the projectors
\begin{equation*}
  P_{\pm} = \frac{1\mp i \gamma_5}{2}.
\end{equation*}
Using $2 \bar{\psi} P_+ \psi = \chi \chi$ we get
\begin{align*}
  ( \Box + m^2) \phi & = - 2 \lambda \bar{\psi} P_- \psi - m \lambda ( \phi \phi + \phi^* \phi + \phi \phi^* ) - \lambda^2 (\phi^* \phi \phi + \phi \phi \phi^*) \\
  (  i \dslash - m) \psi & = \lambda P_+ (\phi \psi + \psi \phi) + \lambda P_- (\phi^* \psi + \psi \phi^*).
\end{align*}

\subsection{The SUSY current}

We first want to discuss the changes that noncommutativity brings in at the classical level. The equations of
motion are the same, we only have to replace the usual product by the noncommutative one. But there are some
changes for the currents.
It is an interesting feature of noncommutative interacting theories that the local\footnote{By \emph{local} we mean expressions that are polynomials of (derivatives) of fields, where the product is the appropriate algebra product, i.e., (\ref{eq:product}) in the present case. Using different products (nonlocal in our sense), it is possible to construct conserved currents, see, e.g., \cite{Dorn, Pengpan}.} currents associated to symmetries are in general not conserved~\cite{EM-Tensor, Diplom}. Examples are the energy--momentum tensor in the $\phi^4$-model~\cite{Micu} and in electrodynamics~\cite{EM-ED}.
Here we show that the local current associated to the supersymmetry transformation is not
conserved in the interacting case, i.e., for $\lambda \neq 0$. We discuss this in terms of the superfield
$\Phi$. The equation of motion is
\begin{equation*}
 - \frac{1}{4} \bar{D}^2 \bar{\Phi} + m \Phi + \lambda \Phi \Phi = 0.
\end{equation*}
The local supercurrent is given by
\begin{equation*}
  V_{\alpha \dot{\alpha}} = \frac{1}{2} [ D_{\alpha} \Phi, \bar{D}_{\dot{\alpha}} \bar{\Phi} ] + i \{ \dslash_{\alpha \dot{\alpha}} \Phi, \bar{\Phi} \} - i \{ \Phi, \dslash_{\alpha \dot{\alpha}} \bar{\Phi} \}.
\end{equation*}
Here we used a symmetrized version of the usual current, since this is usually advantageous in the
noncommutative case.
By standard methods (see, e.g., \cite{Sohnius}) one can show that
\begin{equation*}
  \bar{D}^{\dot{\alpha}} V_{\alpha \dot{\alpha}} = \frac{1}{2} \{ D_{\alpha} \Phi, \bar{D}^2 \bar{\Phi} \} - \frac{1}{4} \{ \Phi, D_{\alpha} \bar{D}^2 \bar{\Phi} \}
\end{equation*}
holds. Using the equation of motion, we get
\begin{align*}
  \bar{D}^{\dot{\alpha}} V_{\alpha \dot{\alpha}} & = 2 \left\{ D_{\alpha} \Phi, \left( m \Phi + \lambda \Phi \Phi \right) \right\} - \left\{ \Phi, D_{\alpha}  \left( m \Phi + \lambda \Phi \Phi \right) \right\} \\
  & = m D_{\alpha} \Phi^2 + \lambda [[ D_{\alpha} \Phi, \Phi], \Phi].
\end{align*}
The first term is already present in the commutative case. It does not affect the charge corresponding to the
supersymmetry transformation, but simply expresses the fact that the theory is not conformal. The second term, however,
is a genuinely noncommutative one. It also affects the SUSY charge. Since it is given by a commutator, the non--conservation of the charge is relevant only at the noncommutativity scale\footnote{Such an effect is to be expected by heuristic considerations~\cite{Doplicher}: Charge conservation requires that the production of a particle with positive charge is always accompanied by the production of a particle with opposite charge at the same place. But because of the noncommutativity, it is not possible to localize two particles at the same place, see, e.g., the discussion in~\cite{UVfinite}.}. Like the non-conservation of the local energy--momentum tensor, this effect does not show up in a perturbative treatment of the corresponding quantum theory, at least not at second order.


\subsection{The self energy}



Now we compute the self energy at the one-loop level.
Using the equations of motion, the first terms in the Yang-Feldman series are
\begin{align}
\label{eq:phi_1}
  \phi_1 & = - \Delta_R \times \left( 2 \bar{\psi}_0 P_- \psi_0 + m ( \phi_0^* \phi_0 + \phi_0 \phi_0^* + \phi_0 \phi_0 )  \right), \\
\label{eq:psi_1}
  \psi_1 & = S_R \times \left( P_+ ( \phi_0 \psi_0 + \psi_0 \phi_0) + P_- ( \phi_0^* \psi_0 + \psi_0 \phi_0^* )
  \right),
\end{align}
and the analogous formulas for the conjugate fields. The second order component of $\phi$ is
\begin{align}
\label{eq:term1}
  \phi_2 = - \Delta_R \times & \Big\{ 2 \bar{\psi}_1 P_- \psi_0 + 2 \bar{\psi}_0 P_- \psi_1 \\
\label{eq:term2}
  & + m ( \phi_1^* \phi_0 + \phi_0^* \phi_1 + \phi_1 \phi_0^* + \phi_0 \phi_1^* + \phi_1 \phi_0 + \phi_0 \phi_1 ) \\
\label{eq:term3}
  & + ( \phi_0 \phi_0 \phi_0^* + \phi_0^* \phi_0 \phi_0 ) \Big\}
\end{align}
Inserting (\ref{eq:phi_1}) and (\ref{eq:psi_1}) in (\ref{eq:term1}) and (\ref{eq:term2}) and contracting the free fields, one can write $\phi_2$ in the form
\begin{equation*}
 \hat \phi_2(k) = (2\pi)^2 \hat \Delta_R(k) \left( \Sigma(k) \hat \phi_0(k) + \Sigma'(k) \hat \phi^*_0(k) \right) + \text{ n.o.},
\end{equation*}
cf. (\ref{eq:phi_2}).


For the computation of the graphs involving fermions, we need the formulae\footnote{The factor $1/2$ in the last
line is due to the Majorana nature of the fermions. 
}
\begin{align*}
  \hat{S}_R(k) & = ( - \kslash - m) \hat{\Delta}_R(k), \\
  \hat{\bar{S}}_R(k) & = ( \kslash - m) \hat{\Delta}_R(k), \\
  \langle \hat{\bar{\psi}}_\alpha(k) \hat{\psi}_\beta(p) \rangle = & \frac{1}{2} (2 \pi)^{2} \delta(k+p) ( - \kslash + m )_{\beta \alpha} \hat{\Delta}_+(k).
\end{align*}

\begin{description}

 \item[The $\phi^4$ tadpole] is obtained from the term (\ref{eq:term3}) of $\phi_2$. We find the quadratically divergent contribution
\begin{equation*}
  \Sigma_{\phi^4-\text{tp}}(k) = - 2 (2\pi)^{-2} \lambda^2 \int \ud^4 l \ \hat{\Delta}_+(l) \left( 1 + e^{i k \sigma l} \right).
\end{equation*}


\item[The $\phi^3$ tadpole] is obtained from the term (\ref{eq:term2}) by contracting the $\phi_0$s in $\phi_1$
or $\phi_1^*$ among themselves. Due to the retarded propagator with zero momentum connecting the loop with the
line, the mass appearing in the interaction term cancels and we get
\begin{equation*}
  \Sigma_{\phi^3-\text{tp}}(k) = 8 (2\pi)^{-2} \lambda^2 \int \ud^4 l \ \hat{\Delta}_+(l).
\end{equation*}
Note that no twisting factor appears.


\item[The $\phi^3$ fish graph] is obtained from the term (\ref{eq:term2}) by contracting a $\phi_0$ in $\phi_1$
or $\phi_1^*$ with the outer $\phi_0^*(f)$. We get
\begin{equation*}
 \Sigma_{\phi^3-\text{fish}}(k) = 3 m^2 \lambda^2 \int \ud^4 l \ \hat{\Delta}_+(l) \left( 1 + e^{i k \sigma l} \right) \left( \hat{\Delta}_R(k-l) + \hat{\Delta}_R(k+l) \right).
\end{equation*}


\item[The Yukawa tadpole] is obtained from (\ref{eq:term2}) by contracting the fermions in $\phi_1$ or
$\phi_1^*$. Since the trace of a single $\gamma$-matrix vanishes we only get a supplementary factor $4 m$ and
thus
\begin{equation*}
  \Sigma_{\text{Yuk}}(k) = - 8 (2\pi)^{-2} \lambda^2 \int \ud^4 l \ \hat{\Delta}_+(l).
\end{equation*}


\item[The fermion fish graph] is obtained from the term (\ref{eq:term1}). The relevant part of $\phi_2$, i.e.,
the part involving $\phi_0$, is
\begin{multline*}
  \hat{\phi}_2(k) = - 4 \hat{\Delta}_R(k) \int \ud^4l \ud^4l' \ \cos \frac{l \sigma l'}{2} \\
 \times \left\{ \hat{\bar{\psi}}_0(k-l) P_- \hat{S}_R(l) P_+ \hat{\psi}_0(l-l') \hat{\phi}_0(l') e^{- \frac{i}{2} k \sigma l} \right. \\
  \left. + \hat{\bar{\psi}}_0(l-l') P_+ \hat{\bar{S}}_R(l) P_- \hat{\psi}_0(k-l) \hat{\phi}_0(l') e^{- \frac{i}{2} l \sigma k} \right\}.
\end{multline*}
Contraction of the fermion fields now yields
\begin{align*}
  & - 2 (2 \pi)^{2} \hat{\Delta}_R(k) \hat \phi_0(k) \int \ud^4l \ \cos \frac{l \sigma k}{2} \\
 & \quad \times \left\{ \tr \left( P_- (-\lslash -m) P_+ (\kslash-\lslash-m) \right) \hat{\Delta}_R(l) \hat{\Delta}_+(k-l) e^{-\frac{i}{2} k \sigma l} \right. \\
 & \qquad \left. + \tr \left( P_+ (\lslash -m) P_- (-\kslash+\lslash-m) \right) \hat{\Delta}_R(l) \hat{\Delta}_+(-k+l) e^{-\frac{i}{2} l \sigma k} \right\} \\
= & - 2 (2 \pi)^{2} \hat{\Delta}_R(k) \hat \phi_0(k) \int \ud^4l \ \cos \frac{l \sigma k}{2} \\
 & \quad \times \left\{ \tr \left( P_- (\lslash - \kslash -m) P_+ (\lslash-m) \right) \hat{\Delta}_R(k-l) \hat{\Delta}_+(l) e^{-\frac{i}{2} l \sigma k} \right. \\
 & \qquad \left. + \tr \left( P_+ (\kslash + \lslash -m) P_- (\lslash-m) \right) \hat{\Delta}_R(k+l) \hat{\Delta}_+(l) e^{-\frac{i}{2} l \sigma k}
 \right\}.
\end{align*}
With the usual $\gamma$ matrix algebra, we get
\begin{multline*}
 \Sigma_{\psi-\text{fish}}(k) = 2 \lambda^2 \int \ud^4 l \ \hat{\Delta}_+(l) \left( 1 + e^{i k \sigma l} \right) \\ \times \left( (k-l) \cdot l  \hat{\Delta}_R(k-l) - (k+l) \cdot l \hat{\Delta}_R(k+l) \right).
\end{multline*}

\end{description}

Now we collect all our terms. The Yukawa tadpole and the $\phi^3$ tadpole cancel (this has to be so in order to
have a vanishing VEV of $\phi_1$). Using
\begin{equation*}
  (l^2 - m^2) \hat{\Delta}_+(l) = 0, \quad (l^2 - m^2) \hat{\Delta}_A(l) = - (2 \pi)^{-2},
\end{equation*}
the combination of the other terms gives
\begin{equation*}
  \Sigma(k) = \lambda^2 \left( k^2 + m^2 \right) \int \ud^4l \ \hat{\Delta}_+(l) \left( 1 + e^{i k \sigma l} \right) \left( \hat{\Delta}_R(k-l) + \hat{\Delta}_R(k+l) \right).
\end{equation*}
Apart from the prefactor $(k^2+m^2)$, this is exactly the expression we already found for the $\phi^3$-model. We remark that for the self--energy of the fermion, one obtains the same result.

The prefactor is to be expected: Assuming that the non-renormalization theorem still holds, we know that only
the $\bar{\Phi} \Phi |_{\theta^2 \bar{\theta}^2}$-term gets renormalized. From the free equations of motion
\begin{equation*}
  (1+\delta Z) F - m \phi^* = 0 , \qquad (1+\delta Z) \Box \phi + m F^* = 0
\end{equation*}
we get, at first order in $\delta Z$,
\begin{equation*}
  (\Box + m^2) \phi = - \delta Z (\Box - m^2) \phi.
\end{equation*}
Note that in our terminology, this corresponds to both a field strength and a mass renormalization. Explicitly,
we have, after subtracting the planar part,
\begin{align}
\label{eq:WZ_M}
  \delta m^2(s) = & - 2 m^2 \Sigma_{np}(m^2,s), \\
\label{eq:WZ_Z}
  \delta Z(s) = & - \Sigma_{np}(m^2,s) - 2 m^2 \frac{\del}{\del k^2} \Sigma_{np}(m^2,s).
\end{align}
Here we used the $\Sigma_{np}$ from the previous section, cf. equation (\ref{eq:F_np}). From (\ref{eq:WZ_M}) we
conclude that for $\sigma=\sigma_0, m=10^{-17} \lambda_{nc}^{-1}, \lambda=1$ the distortion of the group
velocity is twice as strong as in the $\phi^3$--model. Identifying $\phi$ with the Higgs field, 
an effect of this magnitude might
be measurable at the next generation of particle colliders.

As was already discussed in the previous section, the second term in (\ref{eq:WZ_Z}) is effectively constant for
realistic momenta. The first term has already been plotted in Fig.~\ref{fig:M}, apart from the sign. As discussed in Remark~\ref{rem:FSRen}, a momentum--dependent field strength renormalization leads to a nonlocal smearing. In order to estimate its strength, one has to compute the Fourier
transform of $\Sigma_{np}$. In~\cite{InPrep}, such a calculation is performed in the setting of noncommutative supersymmetric electrodynamics.

Note that the mass and field strength renormalizations for the fermion component are exactly the same.

\section{Calculation in the sense of oscillatory integrals}
\label{sec:OscInt}

The aim of this section is to show that~(\ref{eq:F_NP}) is well--defined in the sense of oscillatory integrals, and that a calculation is this sense yields the same result as the formal calculation done in Section~\ref{sec:phi3np}. 
We use the theory of oscillatory integrals as given in~\cite{ReedSimon}. We first state the main definitions and results.

  Let $\Omega$ be an open set in $\R^s$.
  \begin{definition}
    A \emph{phase function} on $\Omega \times \R ^t$ is a continuous function $\phi :\Omega\times \R ^t \to \R$ with
    \begin{enumerate}
      \item $\forall \lambda \ge 0,(k,l)\in \Omega\times \R ^t$: $\phi(k,\lambda l)=\lambda \phi(k,l)$,
      \item $\phi$ is $\cal C^\infty$ on $\Omega\times (\R ^t \backslash \{ 0 \})$,
      \item $(\grad_k\phi,\grad_l \phi) \ne (0,0)$ on $\Omega\times (\R ^t \backslash \{ 0 \})$.
    \end{enumerate}
  \end{definition}

  \begin{definition}
    A $\cal C^\infty$ function $a :\Omega\times \R ^t \to \mathbb{C}$ is called \emph{symbol of order} $r\in \R$ on
    $\Omega\times \R ^t$ if
    $\forall K \subset \Omega$ compact and for all multiindices $\alpha, \beta$ the seminorms
    \begin{equation*}
      \|a\|_{K,\alpha,\beta}=\sup_{k\in K, l \in \R^t} (1+|l|)^{|\beta|-r} |D^\alpha_k D^\beta_l a(k,l)|
    \end{equation*}
    are finite. The set of all such symbols with topology given by the seminorms
    will be denoted by $\Sym(\Omega,t,r)$.

A function $a:\Omega\times \R ^t \to \mathbb{C}$ is called \emph{asymptotic symbol}, if it can be written as $a=a_1+a_2$ with $a_1 \in \Sym(\Omega,t,r)$ and $a_2$ having compact support in $l$ and the map $k\to a_2(k,\cdot)$ is $\cal C^\infty$ as a map from $\Omega$ to $L^\infty(\R^t)$.
\end{definition}

  If $r<r^\prime$ then $\Sym(\Omega,t,r)\subset \Sym(\Omega,t,r^\prime)$ and the $\cal C^\infty$
  functions of compact support
  are dense in $\Sym(\Omega,t,r)$ in the topology of $\Sym(\Omega,t,r^\prime)$.

  For $a_1 \in \Sym(\Omega,t,r_1)$ and $a_2 \in \Sym(\Omega,t,r_2)$ the product $a_1 \cdot a_2 $ is in
  $\Sym(\Omega,t,r_1+r_2)$ and similar for asymptotic symbols.

  Now we want to give a natural extension to expressions like $\int \ud^t l\ a(k,l) e^{i \phi(k,l)}$ if the integral is
  not absolutely convergent:
  \begin{theorem}\label{oszinttheorem}
    Let $\phi$ be a phase function. We can associate with $\phi$ a linear map from the asymptotic symbols
    to ${\cal D} ^\prime(\Omega)$
    denoted by $T_\phi (a)$ and uniquely determined by:
    \begin{enumerate}
      \item If $a$ has compact support in $l$ then $T_\phi (a)(k)=\int \ud^t l\ a(k,l) e^{i \phi(k,l)}$ and is
        a $\cal C^\infty$ function of $k$.
      \item The restriction of $T_\phi$ to $\Sym(\Omega,t,r)$ is a continuous function
        from $\Sym(\Omega,t,r)$ to ${\cal D} ^\prime(\Omega)$.
    \end{enumerate}
  \end{theorem}
  Furthermore, one can show that the singular support of $T_\phi (a)$ is contained in the set
  \begin{equation}\label{eq:singsupp}
  \{k|\exists l \in
  \R^t\backslash\{0\} \text{ with } \grad_l\phi(k,l)=0\}.
  \end{equation}

  \begin{remark}\label{rem:asymptoticsymbol}
  It is easy to see that the notion of asymptotic symbols can be generalized further. The function $a$ could be
  split even further into $a=a_1+a_2+a_3+\ldots$. For the additional terms, $k\to a_i(k,\cdot)$ should again be
  a $\cal C^\infty$ map, having compact support in $l$, into some suitable space of functions or distributions. Example for such spaces would be $L^\infty(\R^t)$, which was already used for the asymptotic symbols, or
  the elements of ${\cal E}'(\R^t)$ which are $\cal C^\infty$ around
  $l=0$.\footnote{As the phase function does not have to be smooth in $l=0$,
  $a_i(k.\cdot)$ should, e.g., not contain derivatives of the $\delta$ function at that point.}
  The important point is that the
  integrals $\int \ud^s k \ f(k) a_i(k,l) e^{i \phi(k,l)}$ should each be well defined for $f \in {\cal D}(\Omega)$,
  one of these in the sense of  oscillatory integrals,
  and their sum independent of the splitting.
  So one could even allow for some $k\to a_i(k,\cdot)$ to be distributions instead of $\cal C^\infty$ maps. This
  could, of course, increase the singular support beyond (\ref{eq:singsupp}).
  \end{remark}

  In our concrete case~(\ref{eq:F_NP}), we choose $\Omega$ to be an open neighbourhood of the mass shell $m$ such that for $k \in \Omega$
  we have $(k\pm l_+)^2 \ne m^2$. For example $\Omega=\{k|\frac m  2< \sqrt{k^2} < \frac {3m} 2 \}$. Furthermore, we have $t=3$, $\phi=-k^\mu \sigma_{\mu \nu} (\betrag{\V l},\V l)^\nu$ and 
  \begin{equation*}
  a(k,l)=  \frac{1}{(2 \pi)^{3}} \frac{1}{2 \omega_\V l}\left(\frac{-1}{(k-l_+)^2-m^2} +\frac{-1}{(k+l_+)^2-m^2}\right)
  e^{-i (k \sigma)_0 (\sqrt{ \V l^2+m^2}-\betrag{\V l})}.
  \end{equation*}
  $a$ is an asymptotic symbol\footnote{It is only asymptotic, since $\betrag{\V l}$ is not
  differentiable at $\V l =0$,
  and one has to use $\sqrt{ \V l^2+m^2}- \betrag{\V l}\le C (1+\betrag{\V l})^{-1}$, cf.~\cite{ReedSimon}.} on $\Omega\times
  \R^3$ of order -3.

  From Theorem \ref{oszinttheorem} we can see that the oscillatory integral
  is a well defined distribution but do not know what
  it looks like. When trying to transform the integral, difficulties arise from the fact that the usual techniques
  of variable transformations are in general not allowed.
  Also the methods used in~\cite{ReedSimon} for the proof of Theorem \ref{oszinttheorem} are not really suitable to make exact or
  numerical calculations.
  Programs for numerical integration can only tackle absolutely convergent or oscillating improper
  Riemann\footnote{An oscillating improper Riemann integral is, e.g., $\lim_{a\to \infty} \int_0^a \ud x \ 1/x \sin x$.} integrals. At
  the end we are going to reduce the oscillatory integral encountered here to an absolutely convergent integral.

  First, the strategy will be to construct an asymptotic symbol with compact support in $\V l$ which approaches $a$
  in the topology of symbols\footnote{We are a little bit sloppy here. To be precise, we would have to write
  $a=a_1+a_2$ like above, using a $\cal C^\infty$ cutoff function around $\V l=0$, and only approximate $a_1$
  by symbols of compact support. It is easy to see that this gives the same result.}
  of some higher order, say, -2. The continuity
  of $T_\phi$ ensures that the result is independent from the way $a$ is approached.

  What we already can deduce is that $T_\phi(a)(k)$ is a $\cal C^\infty$-function of $k$ since
  $\grad_l\phi(k,l)$ is only zero for $k_\mu \sigma^{\mu \nu}$ lightlike and this can never happen on $\Omega$.

  For $k\in \Omega$ let $\Lambda_k$ be the unique pure boost which takes the vector $k$ to $\Lambda_k k=(\sqrt{k^2},\V
  0)$. It is easy to see that $\Lambda_k$ is a $\cal C^\infty$ function of $k$.

  Let $g \in \cal D(\R)$ have the property
  \begin{equation*}
    g(x)=\begin{cases}
           1& \text{ if } \betrag{x} \le 1, \\
           0& \text{ if } \betrag{x} \ge 2.
         \end{cases}
  \end{equation*}

  Define
  \begin{equation*}
    G_n(k,\V l):=g\left((\Vmat{\Lambda}_k l_+/ n)^2\right),
  \end{equation*}
  where $\Vmat{\Lambda}_k$ is only the vector part of the transformation, i.e., a $3 \times 4$ matrix
  and the  square is the Euclidean square of a 3-vector.
  $G_n$ is a $\cal C^\infty$-function of $k$ and $\V l$
  and for given $k,n$ it has compact support in $\V l$ and $\forall n$ lies in $\Sym(\Omega,3,0)$.

  \begin{lemma}
    $G_n\to 1$ in $\Sym(\Omega,3,1)$ for $n \to \infty$.
  \end{lemma}
  \begin{proof}
    We have to show that $\forall K \subset \Omega$ compact and $\forall \alpha,\beta$
    \begin{equation}\label{ggegeneins}
      \sup_{k\in K ,\V l} (1+\betrag{\V l})^{\betrag{\beta}-1} \betrag{ D^\alpha_k D^\beta_{\V l} \left( g \left((\Vmat \Lambda_k l_+/n)^2\right)-1 \right)}
      \xrightarrow[n\to \infty]{} 0.
    \end{equation}

    It is easy to see that $\forall \alpha$
    \begin{equation*}
      \norm{ D^\alpha_k \Vmat \Lambda_k}_{\sup}=:c^\alpha_k
    \end{equation*}
    is a continuous function of $k$ on $\Omega$ and that one can find positive constants $d^\beta$
    such that $\forall \beta$
    \begin{equation*}
      \norm{ D^\beta_{\V l} l_+}_{\text{Euclid}}\le d^\beta(1+ \betrag{\V l})^{1-\betrag{\beta}}.
    \end{equation*}
    With these one can construct $C^{\alpha,\beta}_k$, which are positive continuous functions of $k$, such
    that
    \begin{equation}\label{abschaetzung1}
      \betrag{D^\alpha_k D^\beta_{\V l} ({\Vmat \Lambda_k l_+})^2 } \le C^{\alpha,\beta}_k (1+\betrag{\V l})^{2-\betrag{\beta}}.
    \end{equation}

    First we show (\ref{ggegeneins}) for $\betrag{\alpha}=\betrag{\beta}=0$:
    $\betrag{g \left( (\Vmat \Lambda_k l_+/n)^2\right)-1}$ is only unequal to zero if
    $(\frac {\Vmat \Lambda_k l_+} n)^2\ge 1$. With $C^{0,0}_K:=\sup_{k \in K} C^{0,0}_k$ we then get
    \begin{equation*}
      1+\betrag{\V l}\ge n \frac 1 {\sqrt {C^{0,0}_K}}
    \end{equation*}
    and with this
    \begin{equation*}
       \sup_{k\in K ,\V l \in \R^t} (1+\betrag{\V l})^{-1} \betrag{ g \left( (\Vmat \Lambda_k l_+/n)^2\right)-1}
      \le  \sup_{x \in \R} \betrag{g(x)-1}\sqrt {C^{0,0}_K} \frac 1 n \xrightarrow[n\to \infty]{} 0.
    \end{equation*}

    Now let $\alpha$ or $\beta$ be unequal to zero:
    With (\ref{abschaetzung1}) one can easily see that
    \begin{equation*}
      \betrag{D^\alpha_k D^\beta_{\V l} g \left( (\Vmat \Lambda_k l_+/n)^2 \right)}
      \le \sum_{\gamma=1}^{|\alpha|+|\beta|} \betrag{(\partial^\gamma g) \left( (\Vmat \Lambda_k l_+/n)^2 \right)}
      \frac 1 {n^{2 \gamma}} \tilde C^\gamma_k (1+\betrag{\V l})^{2 \gamma-\betrag{\beta}},
    \end{equation*}
    where $\tilde C^\gamma_k$ are again positive continuous functions of $k$ (and are also depending on
    $\alpha$ and $\beta$). For each $\gamma$ the function $\partial^\gamma g(x)$ is only unequal to 0 if
    $|x|< 2$. It is not hard to prove that one can estimate
    \begin{equation*}
      ({\Vmat \Lambda_k l_+})^2\ge a_k \cdot (1+|\V l |)^2-b_k,
    \end{equation*}
    where $a_k$ and $b_k$ are again positive continuous functions of $k$. If the argument of $g$ is smaller than
    2 it follows
    \begin{equation*}
      \frac {1+|\V l|} n\le \sqrt{\frac{2+\frac {b_k}{n^2}}{a_k}}.
    \end{equation*}
    Now we can deduce
    \begin{align*}
      & \sup_{k\in K ,\V l \in \R^t} (1+\betrag{\V l})^{\betrag{\beta}-1} \betrag{D^\alpha_k D^\beta_{\V l} \left( g\left((\Vmat \Lambda_k l_+/n)^2\right)-1\right) } \\
      \le  & \sup_{k\in K ,\V l \in \R^t} \sum_{\gamma=1}^{\betrag{\alpha+\betrag{\beta}}} \betrag{\partial^\gamma g \left( (\Vmat \Lambda_k l_+ / n)^2 \right) } \tilde C^\gamma_k \frac{(1+ \betrag{\V l})^{2 \gamma-1}}{n^{2 \gamma}}\\
      \le & \sum_{\gamma=1}^{\betrag{\alpha}+\betrag{\beta}}\sup_{x \in \R} \betrag{ \partial^\gamma g(x) } \tilde C^\gamma_K
      \left(\frac{2+\frac {b_K}{n^2}}{a_K}\right)^{\gamma -\frac 1 2} \frac 1 n   \xrightarrow[n\to \infty]{} 0,
    \end{align*}
    with $a_K = \sup_{k \in K} a_k$.
    This completes the proof.
  \end{proof}

  With the above result it follows that $G_n\cdot a$ has compact support in $\V l$ for given $k$
and approaches $a$ in the topology of $\Sym(\Omega,3,-2)$.
  Calculating the integral (\ref{eq:F_NP}), with $f \in {\cal D}(\Omega)$, we get
  \begin{equation}
  \label{eq:F_osz}
     \frac{1}{(2 \pi)^{3}} \int \ud^4 k \frac{\ud^3 l}  {2 \omega_\V l}f(k) g\left( (\Vmat \Lambda_k l_+/n)^2 \right)
     \left(\frac{-1}{(k-l_+)^2-m^2} +\frac{-1}{(k+l_+)^2-m^2}\right)e^{-i k \sigma l_+}.
  \end{equation}
  This integral is absolutely convergent, so the usual techniques for manipulating integrals are allowed. We
  perform a $k$-dependent nonlinear transformation on $\V l$: $\V l^\prime= \Vmat \Lambda_k l_+$. The
  integration measure does not change and, of course, $l_+ = \Lambda_k^{-1} l^\prime_+$.
  The prime will be dropped again and we get:
  \begin{multline}
  \label{eq:Zwischenschritt}
     \frac{1}{(2 \pi)^{3}} \int \ud^4 k f(k) \int   \frac{\ud^3 l}  {2 \omega_\V l}
     \left(\frac{-1}{(k-\Lambda_k^{-1}l_+)^2-m^2} +\frac{-1} {(k+\Lambda_k^{-1}l_+)^2-m^2}\right) \\ \times g\left( (\V l / n )^2 \right)
     e^{-i k \sigma \Lambda_k^{-1}l_+}.
  \end{multline}
  It holds
  \begin{equation*}
    (k\pm\Lambda_k^{-1}l_+)^2=(\Lambda_k^{-1}((\sqrt{k^2},\V 0)\pm l_+))^2=k^2+m^2\pm 2 \omega_{\V l}
    \sqrt{k^2}.
  \end{equation*}
  Thus, the sum of the two fractions in~(\ref{eq:Zwischenschritt}) is
  $\frac{-2}{k^2-4\omega_{\V l}^2} $.
  Define $\sigma' = {\Lambda_k^{-1}}^T \sigma \Lambda_k^{-1}$.
  $\sigma^\prime$ is again antisymmetric, so $(\sqrt{k^2},\V 0)^\mu \sigma^\prime_{\mu \nu}$
  has vanishing time component. Let $\underline{(\sqrt{k^2},\V 0) \sigma^\prime}$ be its spatial part. Its length
  is
\begin{equation*}
  \sqrt{-\left( (\sqrt{k^2},\V 0) \sigma^\prime \right)^2}=\sqrt{-(k \sigma)^2}.
\end{equation*}  
The expression in the exponent in (\ref{eq:Zwischenschritt}) now becomes
  \begin{equation*}
    k \sigma \Lambda_k^{-1}l_+=(\sqrt{k^2},\V 0) \sigma^\prime l_+=- \underline{(\sqrt{k^2},\V 0) \sigma^\prime}\cdot \V l.
  \end{equation*}
  We use spherical coordinates for $\V l$ where the $z-$axis is along $\Vmat{(\sqrt{k^2},\V 0) \sigma^\prime}$.
  The exponent equals $\sqrt{-(k \sigma)^2} l \cos(\theta)$, and after performing the $\phi$ and $\theta$
  integration we get (dropping the $k$-integration)
  \begin{equation*}
    -2  (2 \pi)^{-2} \int_0^\infty\ud l \ g\left( (l / n)^2 \right) \frac {l^2}{\omega_l(k^2-4\omega_l^2)}
     \frac {\sin(l \sqrt{-(k \sigma)^2})}{l \sqrt{-(k \sigma)^2}}.
  \end{equation*}
  For $n \to \infty$ this gives the value of $T_\phi(a)(k)$, which is the absolutely convergent integral
  \begin{equation*}
   -2  (2 \pi)^{-2} \int_0^\infty\ud l \ \frac {l^2}{\omega_l(k^2-4\omega_l^2)}
     \frac {\sin(l \sqrt{-(k \sigma)^2})}{l \sqrt{-(k \sigma)^2}},
  \end{equation*}
  which is the same result as (\ref{eq:F_np}).

  We emphasize again that in order to calculate the dispersion relation at the one-loop level, it is sufficient to know
  \begin{equation}
  \Sigma_{np}(k) = \int \ud^4l \ \hat{\Delta}_+(l) e^{i k \sigma l} \left( \hat{\Delta}_{R}(k-l) + \hat{\Delta}_{R}(k+l)
  \right), \tag{copy of \ref{eq:F_NP}}
  \end{equation}
  for $k$ in the vicinity of the mass shell. However, when it comes to treat higher orders, the fish-graphs, which
  give the contributions (\ref{eq:F_NP}), may appear as subgraphs and have to be integrated over arbitrary $k$.
  Then the problem appears that $\hat{\Delta}_R(k \pm l_+)$ can become singular, so that~(\ref{eq:F_NP}) is no
  oscillatory integral in the standard sense. Let us examine this more closely: For $k^2 > 4 m^2$, the singular
  support of $\V{l} \to \hat{\Delta}_R(k \pm l_+)$ is compact and does not contain the origin. We may then
  proceed as indicated in Remark \ref{rem:asymptoticsymbol}. Let $k_0 > 0$. Then only $\hat{\Delta}_R(k - l_+)$ can become singular
  and at the singularity we have $k_0 - \omega_l > 0$. Thus, we may simply add $\pm i \epsilon$ to the denominator
  of the first fraction in~(\ref{eq:F_osz}). Of course one then has to assume that $f$ has compact support in $\{
  k \in \R^4 | k^2> 4 m^2, k_0 > 0 \}$. One can then proceed as above and obtains~(\ref{eq:F_np}), but with $(k^2
  - 4 \omega_l^2 \pm i \epsilon)$ in the denominator. Using $\frac{1}{x \pm i \epsilon} = \pv{x} - i \pi
  \delta(x)$, this can be split into real and imaginary part. The imaginary part resembles the usual imaginary
  parts for forward/backward scattering.

For spacelike $k$, the singular support of $\V{l} \to \hat{\Delta}_R(k \pm l_+)$ is not compact. Consider, e.g., $k = (0,0,0,k_z)$. Then $\hat{\Delta}_R(k - l_+)$ is singular on the hyperplane $l_3= 2 k_z$.
Thus, it is not possible to use the framework indicated in Remark \ref{rem:asymptoticsymbol}. One has to extend
the framework further in order to accommodate for symbols whose singularities are not compactly supported.
  There are two natural Ans\"atze for such an extension:
  \begin{enumerate}
  \item The distributions $a$ could be approximated by a sequence of symbols $(a_n)_{n\in \mathbb N}$. For each $a_n$ the
  oscillatory integral is well defined. The oscillatory integral for $a$ can then be achieved if one calculates the
  limit $n\to \infty$ after integrating, if this is well defined and to a large extent independent of the choice
  of the sequence.
  \item One could see the relation
    \begin{equation}\label{eq:extension}
      \int \ud^s k \ud^t l \ f(k) a(k,l) e^{i \phi(k,l)}=
      \lim_{n\to \infty}\int \ud^s k \ud^t l \ f(k) g_n(l) a(k,l) e^{i \phi(k,l)}
    \end{equation}
    for a sequence $g_n$ of symbols with compact support and approaching $1$, as a definition. The right hand side
    of (\ref{eq:extension}), with finite $n$, is even defined for $a$ being some distribution. If the limit
    exists and is independent of the choice of the sequence $g_n$ out of some large class of sequences, this
    would be a reasonable extension.
  \end{enumerate}

We would also like to mention the approach followed in~\cite{InPrep}: There, the nonplanar loop integral is interpreted as a function $F(k,y)$ of two independent variables $k$ and $y$, where the twisting factor is written as $e^{-i y l_+}$. One can show that the integral is a well--defined tempered distribution in $\R^8$. The question is then if it is possible to restrict $y$ to $k \sigma$. Whether the loop integral is well-defined is then a question that can be answered by computing $F(k,y)$. The problem is that it is rather difficult to perform such a calculation analytically.

\begin{remark}
The nonplanar loop integrals that appear in the setting of the modified Feynman rules can also be treated rigorously in the sense of oscillatory integrals. Since one is working in the Euclidean metric there, the symbols can not become singular, so that there are no problems for spacelike external momenta. However, as already mentioned in the introduction, it is not clear whether there is any relation between the results for Euclidean and Minkowski metric.
\end{remark}

\section{Summary and Outlook}

We discussed dispersion relations in the Yang--Feldman formalism at the one-loop level and computed them in the noncommutative $\phi^3$ and Wess--Zumino model. It turned out that the distortions of the dispersion relation
were moderate for parameters typically expected for the Higgs field.
We also showed that the local SUSY current is not conserved in the noncommutative Wess-Zumino model.


  A shortcoming of the present work is of course the lack of a systematic treatment of renormalizability. In the case of the noncommutative Euclidean space, it is usually argued
  that the IR-divergence induced by the UV--IR mixing can at most be of the same degree as the
  underlying UV--divergence, i.e., logarithmic in the two cases studied here. Then the integration over a non--planar
  subgraph poses no problem. However, in the present situation of the noncommutative Minkowski space we have the difficulties mentioned at the end of Section~\ref{sec:OscInt}. To solve these, an
  extension of the mathematical framework of oscillatory integrals is needed.


\begin{center}
{\bf Acknowledgements}
\end{center}

  We would like to thank Klaus Fredenhagen for valuable comments and discussions.  Financial support from the
  Graduiertenkolleg ``Zuk\"unftige Entwicklungen in der Teilchenphysik'' is gratefully acknowledged. Part of this
  work was done while J.~Z. visited the Dipartimento di Matematica of the Universit\`a di Roma ``La Sapienza''
  with a grant of the research training network ``Quantum Spaces -- Noncommutative Geometry''. It is a pleasure to
  thank Sergio Doplicher for kind hospitality.


\begin{thebibliography}{99}

\bibitem{DFR}
S.~Doplicher, K.~Fredenhagen and J.~E.~Roberts, ``The Quantum structure of space-time at the Planck scale and
quantum fields,'' Commun.\ Math.\ Phys.\  {\bf 172} (1995) 187 [arXiv:hep-th/0303037].

\bibitem{SW}
N.~Seiberg and E.~Witten, ``String theory and noncommutative geometry,'' JHEP {\bf 9909} (1999) 032
[arXiv:hep-th/9908142].

\bibitem{EField}
N.~Seiberg, L.~Susskind and N.~Toumbas, ``Strings in background electric field, space/time noncommutativity  and
a new noncritical string theory,'' JHEP {\bf 0006} (2000) 021 [arXiv:hep-th/0005040].

\bibitem{Filk}
T.~Filk, ``Divergencies in a field theory on quantum space,'' Phys.\ Lett.\ B {\bf 376} (1996) 53.

\bibitem{DorosDiss}
D.~Bahns, ``Perturbative methods on the noncommutative Minkowski space,'' DESY-THESIS-2004-004

\bibitem{NCPertDyn}
S.~Minwalla, M.~Van Raamsdonk and N.~Seiberg, ``Noncommutative perturbative dynamics,'' JHEP {\bf 0002} (2000)
020 [arXiv:hep-th/9912072].

\bibitem{GomisMehen}
J.~Gomis and T.~Mehen, ``Space-time noncommutative field theories and unitarity,'' Nucl.\ Phys.\ B {\bf 591}
(2000) 265 [arXiv:hep-th/0005129].


\bibitem{Sibold}
Y.~Liao and K.~Sibold, ``Time-ordered perturbation theory on noncommutative spacetime: Basic  rules,'' Eur.\
Phys.\ J.\ C {\bf 25} (2002) 469 [arXiv:hep-th/0205269].

\bibitem{UVfinite}
D.~Bahns, S.~Doplicher, K.~Fredenhagen and G.~Piacitelli, ``Ultraviolet finite quantum field theory on quantum
spacetime,'' Commun.\ Math.\ Phys.\  {\bf 237} (2003) 221 [arXiv:hep-th/0301100].

\bibitem{DoroAvHam}
D.~Bahns, ``Ultraviolet finiteness of the averaged Hamiltonian on the  noncommutative Minkowski space,''
arXiv:hep-th/0405224.

\bibitem{Heslop}
P.~Heslop and K.~Sibold, ``Quantized equations of motion in non-commutative theories,'' Eur.\ Phys.\ J.\ C {\bf
41} (2005) 545 [arXiv:hep-th/0411161].

\bibitem{Ohl}
T.~Ohl, R.~R\"uckl and J.~Zeiner, ``Unitarity of time-like noncommutative gauge theories: The violation of  Ward
identities in time-ordered perturbation theory,'' Nucl.\ Phys.\ B {\bf 676} (2004) 229 [arXiv:hep-th/0309021].

\bibitem{Langmann}
E.~Langmann and R.~J.~Szabo, ``Duality in scalar field theory on noncommutative phase spaces,'' Phys.\ Lett.\ B
{\bf 533} (2002) 168 [arXiv:hep-th/0202039].

\bibitem{GrosseWulkenhaar}
H.~Grosse and R.~Wulkenhaar, ``Renormalisation of $\phi^4$-theory on non-commutative $\R^4$ to all orders,''
Lett.\ Math.\ Phys.\  {\bf 71} (2005) 13 [arXiv:hep-th/0403232].

\bibitem{YF}
C.~N.~Yang and D.~Feldman, ``The S Matrix In The Heisenberg Representation,'' Phys.\ Rev.\  {\bf 79} (1950) 972.

\bibitem{Moller}
C.~M{\o}ller,
``On the problem of convergence in non-local field theories,''
and the following talks and discussions in the
Proceedings of the International Conference of Theoretical Physics 1953,
Science Council of Japan, 1954.

\bibitem{Marnelius}
R.~Marnelius,
``Can The S Matrix Be Defined In Relativistic Quantum Field Theories With
Nonlocal Interaction?,''
Phys.\ Rev.\ D {\bf 10} (1974) 3411.

\bibitem{BDFP}
D.~Bahns, S.~Doplicher, K.~Fredenhagen and G.~Piacitelli, ``On the unitarity problem in space/time
noncommutative theories,'' Phys.\ Lett.\ B {\bf 533} (2002) 178 [arXiv:hep-th/0201222].

\bibitem{Quasiplanar}
D.~Bahns, S.~Doplicher, K.~Fredenhagen and G.~Piacitelli, ``Field theory on noncommutative spacetimes:
Quasiplanar Wick products,'' Phys.\ Rev.\ D {\bf 71} (2005) 025022 [arXiv:hep-th/0408204].

\bibitem{ReedSimon}
M.~Reed and B.~Simon, ``Methods of Modern Mathematical Physics II: Fourier Analysis, Self-Adjointness,''
Academic Press 1975.

\bibitem{WulkenhaarLorentz}
A.~A.~Bichl, J.~M.~Grimstrup, H.~Grosse, E.~Kraus, L.~Popp, M.~Schweda and R.~Wulkenhaar, ``Noncommutative
Lorentz symmetry and the origin of the Seiberg-Witten  map,'' Eur.\ Phys.\ J.\ C {\bf 24} (2002) 165
[arXiv:hep-th/0108045].

\bibitem{Twist}
R.~Oeckl, ``Untwisting noncommutative $\R^d$ and the equivalence of quantum field theories,'' Nucl.\ Phys.\ B
{\bf 581} (2000) 559
[arXiv:hep-th/0003018]. \\
M.~Chaichian, P.~P.~Kulish, K.~Nishijima and A.~Tureanu, ``On a Lorentz-invariant interpretation of
noncommutative space-time and  its implications on noncommutative QFT,'' Phys.\ Lett.\ B {\bf 604} (2004) 98
[arXiv:hep-th/0408069]. \\
J.~Wess, ``Deformed coordinate spaces: Derivatives,'' arXiv:hep-th/0408080. \\
J.~Zahn, ``Remarks on twisted noncommutative quantum field theory,'' Phys.\ Rev.\ D {\bf 73} (2006) 105005
[arXiv:hep-th/0603231].


\bibitem{Raamsdonk}
M.~Van Raamsdonk and N.~Seiberg, ``Comments on noncommutative perturbative dynamics,'' JHEP {\bf 0003} (2000)
035 [arXiv:hep-th/0002186].

\bibitem{Grosse}
H.~Grosse and H.~Steinacker, ``A nontrivial solvable noncommutative $\phi^3$ model in 4 dimensions,''
arXiv:hep-th/0603052.

\bibitem{Rivelles}
H.~O.~Girotti, M.~Gomes, V.~O.~Rivelles and A.~J.~da Silva, ``A consistent noncommutative field theory: The
Wess-Zumino model,'' Nucl.\ Phys.\ B {\bf 587} (2000) 299 [arXiv:hep-th/0005272].


\bibitem{AdLim}
C.~D\"oscher and J.~Zahn, ``Infrared cutoffs and the adiabatic limit in noncommutative spacetime,'' Phys.\ Rev.\
D {\bf 73} (2006) 045024 [arXiv:hep-th/0512028].

\bibitem{LiaoSibold}
Y.~Liao and K.~Sibold,
``Spectral representation and dispersion relations in field theory on noncommutative space,''
Phys.\ Lett.\ B {\bf 549} (2002) 352
[arXiv:hep-th/0209221].



\bibitem{SUSYNCMink}
C.~S.~Chu and F.~Zamora, ``Manifest supersymmetry in non-commutative geometry,'' JHEP {\bf 0002} (2000) 022
[arXiv:hep-th/9912153]. \\
S.~Ferrara and M.~A.~Lledo, ``Some aspects of deformations of supersymmetric field theories,'' JHEP {\bf 0005}
(2000) 008 [arXiv:hep-th/0002084].

\bibitem{UVIRNote}
I.~Y.~Aref'eva, D.~M.~Belov and A.~S.~Koshelev, ``A note on UV/IR for noncommutative complex scalar field,''
arXiv:hep-th/0001215.

\bibitem{WessBagger}
J.~Wess and J.~Bagger, ``Supersymmetry and supergravity,'' Princeton University Press 1992.

\bibitem{EM-Tensor}
A.~Gerhold, J.~Grimstrup, H.~Grosse, L.~Popp, M.~Schweda and R.~Wulkenhaar,
``The energy-momentum tensor on noncommutative spaces: Some pedagogical
comments,''
arXiv:hep-th/0012112.

\bibitem{Diplom}
J.~Zahn,
``Action and locality principle for noncommutative scalar field theories.  (In
German),'' Diplomarbeit, Universit\"at Hamburg,
DESY-THESIS-2003-041.

\bibitem{Micu}
A.~Micu and M.~M.~Sheikh Jabbari,
``Noncommutative $\phi^4$ theory at two loops,''
JHEP {\bf 0101} (2001) 025
[arXiv:hep-th/0008057].

\bibitem{Dorn}
M.~Abou-Zeid and H.~Dorn,
``Comments on the energy-momentum tensor in non-commutative field theories,''
Phys.\ Lett.\ B {\bf 514} (2001) 183
[arXiv:hep-th/0104244].

\bibitem{Pengpan}
T.~Pengpan and X.~Xiong,
``A note on the non-commutative Wess-Zumino model,''
Phys.\ Rev.\ D {\bf 63} (2001) 085012
[arXiv:hep-th/0009070].

\bibitem{EM-ED}
J.~M.~Grimstrup, B.~Kloibock, L.~Popp, V.~Putz, M.~Schweda and M.~Wickenhauser,
``The energy-momentum tensor in noncommutative gauge field models,''
Int.\ J.\ Mod.\ Phys.\ A {\bf 19} (2004) 5615
[arXiv:hep-th/0210288].

\bibitem{Doplicher}
S.~Doplicher,
private communication.

\bibitem{Sohnius}
M.~F.~Sohnius, ``Introducing Supersymmetry,'' Phys.\ Rept.\  {\bf 128} (1985) 39.

\bibitem{InPrep}
J.~Zahn, ``Dispersion relations in quantum electrodynamics on the noncommutative Minkowski space,'' PhD thesis, Universit\"at Hamburg, DESY-THESIS-2006-037.

\bibitem{Claus}
C.~D\"oscher, ``Yang-Feldman formalism on the noncommutative Minkowski space,'' PhD thesis, Universit\"at Hamburg, DESY-THESIS-2006-032.

\end{thebibliography}
\end{document}